\documentclass{sig-alternate-10pt}

\usepackage{epsf}
\usepackage{cite}
\usepackage{amssymb}
\usepackage{subfigure}
\usepackage{graphicx}
\usepackage[ruled,vlined]{algorithm2e}
\usepackage{float}
\usepackage[T1]{fontenc}
\usepackage{url}

\begin{document}

\title{Dual-structure Data Center Multicast Using Software Defined Networking}

\numberofauthors{2}
\author{
\alignauthor
Wenzhi Cui\\
       \affaddr{Software Institute}\\
       \affaddr{Nanjing University, China}\\
       \email{cwz10@software.nju.edu.cn}
\alignauthor
Chen Qian\\
       \affaddr{Department of Computer Science}\\
       \affaddr{University of Kentucky, KY}\\
       \email{qian@cs.uky.edu}
}

\maketitle
\begin{abstract}
Data center applications use multicast as an effective method to reduce bandwidth cost. However, traditional multicast protocols designed for IP networks are usually bottlenecked by the limited state capacity on switches.  In this paper, we propose a scalable multicast solution on fat tree networks based on the observation that data center multicast traffic has strong heterogeneity. We propose to remove the multicast management logic from switches and use the SDN controller to manage multicast groups. The proposed Dual-structure Multicast (DuSM) determines elephant and mice groups according to their traffic amounts and treats them separately. For each elephant group, the controller installs multicast state to maintain multiple shared trees and the group traffic will be balanced evenly among the trees to avoid congestion. For mice groups, the controller applies state-free mutlicast that trades bandwidth capacity for state capacity, such as multicast-to-unicast translation. Our experiments using real multicast traffic data show that the number of groups DuSM supports can be 300\% of that of IP multicast. DuSM also achieves traffic balance among links.

\end{abstract}



\section{Introduction}
Many Data center applications, such as publish-subscribe services used in cloud computing \cite{Jokela:2009:LLS:1594977.1592592}, data backup and updates, and network virtualization \cite{VXLAN}, rely on multicast to disseminate data to multiple receivers \cite{Vigfusson:2010:DMR:1755913.1755949}. Multicast benefits the network by reducing bandwidth overhead and latency between group members.

Traditional IP multicast protocols cannot scale in Data Center Networks (DCNs) because switches only support limited multicast state in
their forwarding tables \cite{Jokela:2009:LLS:1594977.1592592, Vigfusson:2010:DMR:1755913.1755949, DanLi2011ESM, Cao:2012:DSE:2413176.2413182, Li:2013:SIM:2535372.2535380}.
End-host multicast \cite{Chu:2000:CES:339331.339337, Banerjee:2002:SAL:633025.633045}, also called application-layer multicast, does not store multicast state on routers or switches. However using end-host multicast in DCNs will cause significant bandwidth loss due to large amount of duplicated packets transmitted in the network. Furthermore, most multicast protocols do not utilize the topology 
of DCNs, e.g., multiple parallel paths connecting any pair of end hosts. For example, multicast protocols similar to the Protocol Independent Multicast Sparse Mode (PIM-SM) \cite{PIMSM} usually chooses a single core switch on a Fat Tree as the rendezvous point (RP), i.e., the root of the multicast tree. When groups using the same core switch have traffic bursts at the same time, network congestion may occur.

In this paper, we focus on developing a scalable multicast protocol which can scale multicast in DCNs and balance traffic to reduce congestion.  We start our design by analyzing real multicast traffic traces in DCNs. Results show that group traffic in DCNs has strong heterogeneity. A small fraction of groups contribute to the majority of traffic, called \emph{elephant groups}, while most groups have very low traffic volume, called \emph{mice groups}. Motivated by such observation, we propose a Dual-Structure Multicast (DuSM) system which relies on the controller in a software defined networking (SDN) platform \cite{ONF} to manage multicast groups.
The SDN controller collects network information 
and categorize multicast groups into the two types. For an elephant group, DuSM installs multicast rules on switches to maintain multiple shared trees. Traffic of this group will be balanced among these trees to avoid congestion. For mice groups, which are majority, DuSM applies state-free multicast that trades bandwidth capacity for state capacity, such as multicast-to-unicast translation, and hence saves large multicast state space.

We have implemented our protocol on the \textit{ns-2} Simulator and evaluated our architecture using multicast traffic data from real DCNs. Experimental results show that DuSM can enhance multicast state capacity and achieve  traffic balance among links.

The rest of this paper is organized as follows. Section \ref{sec:background} introduces background knowledge and observations from data analysis. Section \ref{sec:design} presents the algorithm design of our multicast system DuSM. We evaluate the performance of DuSM in Section \ref{sec:simulation} and discuss some open issues and future work in Section \ref{sec:challenges}. We present related work in Section \ref{sec:related}. Finally we conclude our work in Section \ref{sec:conclusion}.

\section{Background and Data Analysis}
\label{sec:background}

\begin{figure}[t]
\centering
\includegraphics[width=7.5cm]{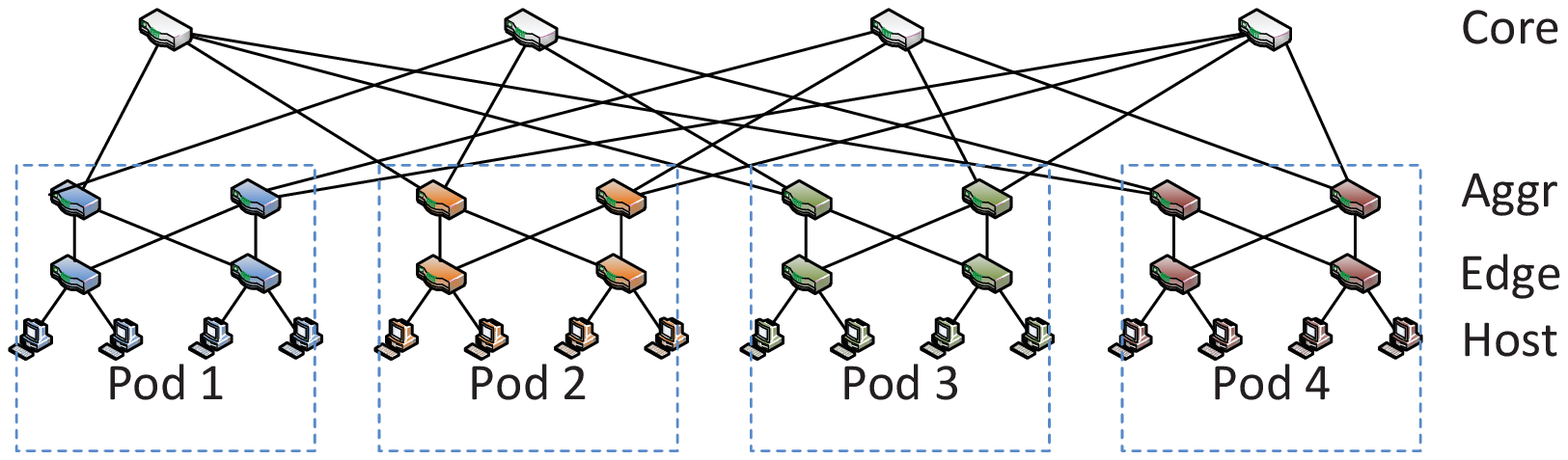}
\vspace{-3ex}
\caption{A Fat Tree topology for a data center network}
\label{fig:fattree}
\vspace{-4ex}
\end{figure}

\begin{figure*}[t]
\begin{center}
\begin{tabular}{p{200pt}p{200pt}}
\subfigure[Histogram of WVE]
 {
   \includegraphics[width=5.5cm]{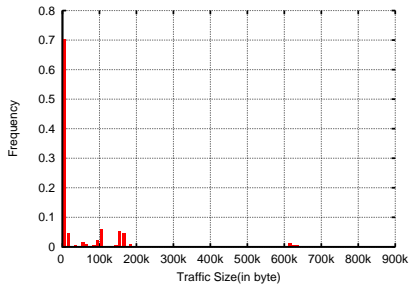}
   \label{fig:tfc-wve-freq}
 }
 &
\subfigure[Histogram of Telecom]
 {
   \includegraphics[width=5.5cm]{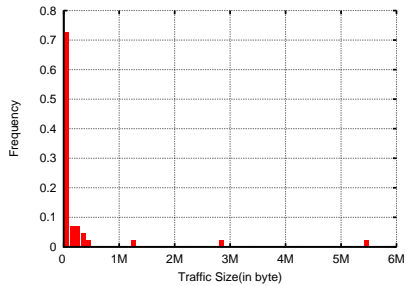}
   \label{fig:tfc-att-freq}
 }
\end{tabular}
\vspace{-3ex} \caption{Group traffic distribution in real data center networks} \label{fig:traffic_dist}
\end{center}
\vspace{-4ex}
\end{figure*}

\subsection{Data Center Topologies}
Today's data center networks often use \emph{multi-rooted hierarchical tree} topologies (e.g., fat tree \cite{fattree} and Clos \cite{clos}) to exploit multiple parallel paths between any pair of hosts. A standard fat tree topology has three vertical layers: edge, aggregate, and core. A \emph{pod} is a management unit down from the core layer, which consists of a set of interconnected end hosts and a set of edge and aggregate switches that connect these hosts. As illustrated in Figure \ref{fig:fattree}, a \emph{fat-tree network} is built from a large number of $k$-port switches and end hosts. There are $k$ pods, interconnected by $(k/2)^2$ core switches. Every pod consists of $(k/2)$ edge switches and $(k/2)$ aggregate switches. Each edge switch also connects $(k/2)$ end hosts. In the example of Figure \ref{fig:fattree}, $k=4$. The current design of DuSM focuses on a fat tree topology. However, we believe our design can be easily extended to other hierarchical topologies, and the design on arbitrary topologies is for future work.

\subsection{Group Traffic Distribution}

We analyze real data center multicast traffic data from two sources, IBM WebSphere Virtual Enterprise (WVE) \cite{IBMWebSphere} that is widely used for other multicast design \cite{Vigfusson:2010:DMR:1755913.1755949, Li:2013:SIM:2535372.2535380} and another data center network of a large telecommunication corporation (we use Telecom as the name of this trace). Due to privacy concerns, all traffics are samples by an unknown rate in each set of data. Hence we only focus on the traffic distribution and pay less attention to the absolute traffic volumes.   The WVE trace has 127 hosts in 1364 groups and the Telecom trace has 512 Virtual Machines (VMs) in 44 groups. For each group, we calculated the total amount of traffic samples in bytes.

Figure~\ref{fig:tfc-wve-freq} and Figure~\ref{fig:tfc-att-freq} show the traffic distribution of the two multicast traces in frequency histograms. We find similar observations from the two sets of data. Although the maximum traffic size in both traces reach a large value, around 70\% groups only have extremely small multicast traffic. A small proportion of groups contribute to the majority of traffic.  We find  strong heterogeneity of multicast traffic in both DCNs. Our observation also validates the assumption made by other multicast design that multicast group traffic follows a power law distribution  \cite{DanLi2011ESM}.

\subsection{Multicast Management by SDN}
\label{sec:controller}

A traditional multicast solution uses a decentralized protocol like IGMP \cite{rfc3228} for group management. As a consequence, multicast routing and group management are usually bundled by listening to IGMP messages and maintaining a multicast forwarding table accordingly. However, emerging trends in SDN provide opportunities to move group management logic from switches or routers to the controller. Recent work has suggested to use the SDN controller for multicast management such as group join and leave \cite{Li:2013:SIM:2535372.2535380, Domainflow}. Our unique contribution is to let the SDN controller utilizes the traffic heterogeneity property and data center topologies for scalable multicast system design.

Centralized group controller can manage multicast group membership in different ways: for example the controller can intercept IGMP messages at edge switches or VM hypervisors for backward compatibility. Although in this work we consider the controller as centralized, DuSM can also deploy multiple controllers in a distributed manner for load balancing and fault tolerance purposes. For example, we can partition the multicast address space into multiple blocks and assign each controller different blocks. The SDN controller can also keep track of group traffic volumes and dynamics using measurement tools on switches \cite{opensketch} so that appropriate actions can be applied for different groups.
Our protocol uses features that can be implemented in most commodity switches, like ECMP routing\cite{rfc6754}, multicast forwarding, IP over IP tunnelling \cite{rfc1853} and GRE encapsulation \cite{rfc1701}. Furthermore, we have a discussion on how to implement the protocol on OpenFlow-enabled switches in Section~\ref{sec:openflow_discussion}.


\section{DuSM Design}
\label{sec:design}

\subsection{Group Classification}
The essential idea of DuSM is to use state-free multicast for the majority mice groups that contribute very little traffic, and save the state capacity for elephant groups. In this paper, a multicast group is identified by a unique class D IP address, which is compatible with current sparse-mode multicast protocols like PIM-SM \cite{PIMSM}. Motivated by our observation on group traffic distribution, the controller first classifies multicast groups based on their traffic amounts and apply different strategies to elephant and mice groups. In DuSM, elephants are large groups whose traffic amount is higher than a threshold. The other groups are called mice groups. As discussed in Section~\ref{sec:controller}, most groups would be classifies as mice if the threshold is set appropriately.

The group classification module of the controller keeps track of group traffic by periodically collecting flow statistics from switches using existing measurement tools \cite{opensketch}. All new groups are treated as mice. When the traffic of a group traffic exceeds a pre-defined threshold, this group will be promoted to elephant and new multicast rules will be installed on corresponding switches.

\subsection{Multicast for Mice Groups}
\label{sec:branching}
DuSM applies state-free multicast to mice groups by using techniques that trade bandwidth capacity for state capacity. We here provide two possibilities.

(1) \emph{Multicast to unicast translation.} We allow the hypervisor of a server obtains information of all mice groups in which its VMs participate, by consulting the controller.  When a hypervisor receives a multicast packet from one of its VMs, it produces multiple duplicate packets, each of which is sent to a single receiver. Mice group packets will then be forwarded by unicast forwarding rules on switches. When the hypervisor of a receiver gets the packet, it can translate the packet back to a multicast packet for backward compatibility. There are many existing techniques to translate a multicast header to unicast-like packet header by rewriting or overloading socket operations \cite{Vigfusson:2010:DMR:1755913.1755949}, IP over IP tunnelling \cite{rfc1853}, or Generic Routing Encapsulation (GRE) \cite{rfc1701} to encapsulate multicast packets in the network layer. We use this solution in our experiments.

(2) \emph{In-packet Bloom filter} has been proposed for state-free multicast \cite{Jokela:2009:LLS:1594977.1592592} \cite{DanLi2011ESM}. Extra processing time and bandwidth may be needed to forward multicast packets with in-packet Bloom filter.  



\subsection{Multicast for Elephant Groups}
When the controller detects that the traffic of a group exceeds a pre-defined threshold, it treats the group as an elephant group.  If all group members are in a same pod, the controller uses a single shared tree for the group. Otherwise, the controller constructs multiple shared trees for the group. Given a core switch as the RP we can uniquely and trivially determine a Steiner tree for a group on a fat tree. Hence the controller  randomly selects multiple cores and computes a Steiner tree for each of them. The controller then installs multicast rules on the switches of these trees. A packet will be matched to one of the trees and delivered along the tree to all receivers. We expect that all packets of the group can be matched evenly on these trees to balance bandwidth cost. Note that for TCP-based unicast, splitting packets on different paths is undesired because it will cause out-of-order delivery. However most multicast applications do not use TCP as the transport protocol, thus in-order delivery is not required.

For example, the controller computes four Steiner trees for an elephant group. The edge switch of a group member would store four rules for these trees. When the edge switch receives a multicast packet, it matches the packet to one Steiner tree. The matching method is static, i.e., a packet will always be matched to a particular Steiner tree on different switches. A possible implementation could be hashing a small piece of packet-related data (such as the checksum) to four bins, each of which corresponds to a tree. The packet will then be forwarded along the tree to all receivers.

\subsection{Discussion on OpenFlow Implementation}
\label{sec:openflow_discussion}
In this section we will discuss possible approaches to implement our protocol on an OpenFlow-enabled platform.
An OpenFlow switch usually has one or more flow tables.
Each flow table contains a set of flow entries with matching fields, count and forwarding actions. For switches supporting OpenFlow 1.1 \cite{OpenFlow-spec-v1.1.0} or higher, actions in the flow table can direct packets to a group table for advanced forwarding actions like multicast. Multicast rules representing the Steiner trees computed by the SDN controller are installed in the group table. All group management requests such as group join and leave are directed to the controller, and the controller can update multicast rules accordingly.
The packet translation module for mice groups is installed to the hypervisors on physical servers. For each receiver of the group, this module simply creates a duplicate unicast packet and write the receiver address onto the destination field.  A simpler way is to apply existing protocols such as IP over IP tunnelling and GRE. In fact, GRE/L3 tunnelling has been listed in OpenFlow 1.2 proposals \cite{OpenFlow-1.2-proposals}. This packet will then be forwarded by unicast rules of switches. Since unicast rules can always be aggregated to wildcard rules \cite{fattree}, they usually do not have scalability problem.

\begin{figure*}[t]
\begin{center}
\begin{tabular}{p{150pt}p{150pt}p{150pt}}
\subfigure[Edge state  overhead]
 {
   \includegraphics[width=5.7cm]{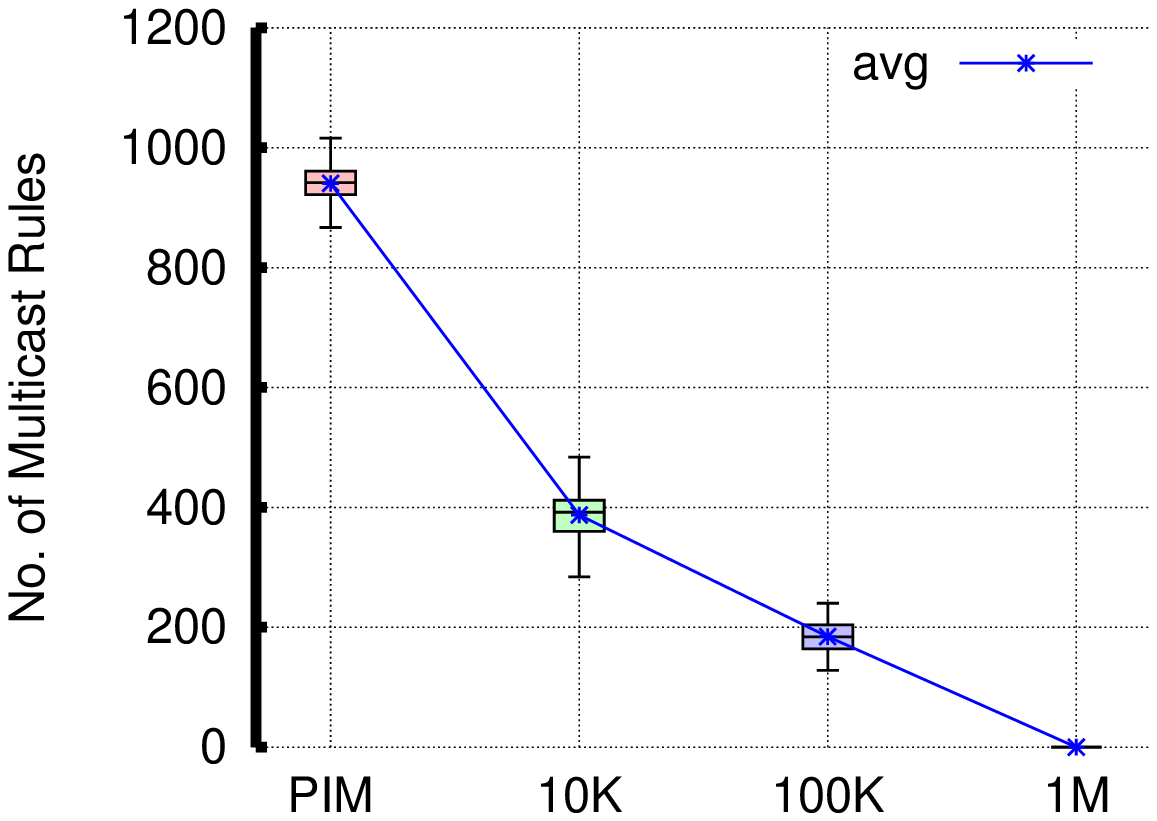}
   \label{fig:wve1024-16k-random-states-edges}
 }
 &
\subfigure[Aggr state  overhead]
 {
   \includegraphics[width=5.7cm]{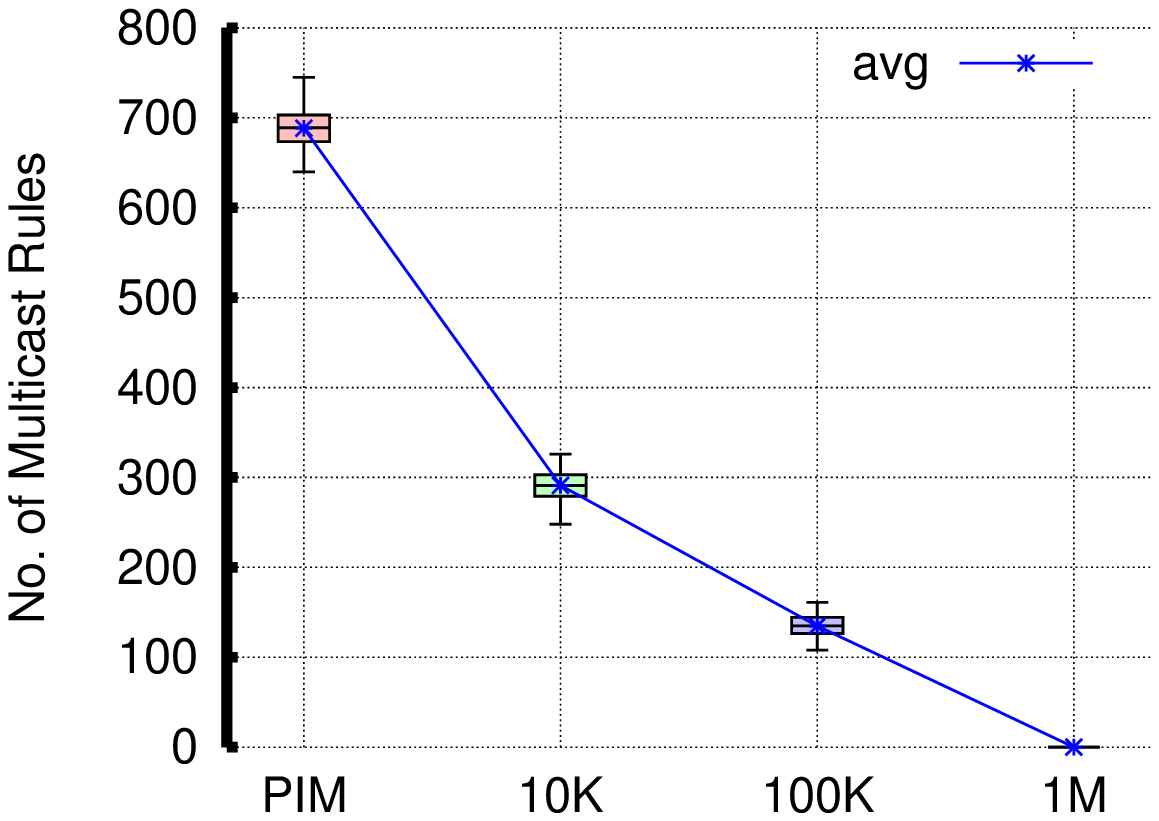}
   \label{fig:wve1024-16k-random-states-aggrs}
 }
 &
\subfigure[Core state  overhead]
 {
   \includegraphics[width=5.7cm]{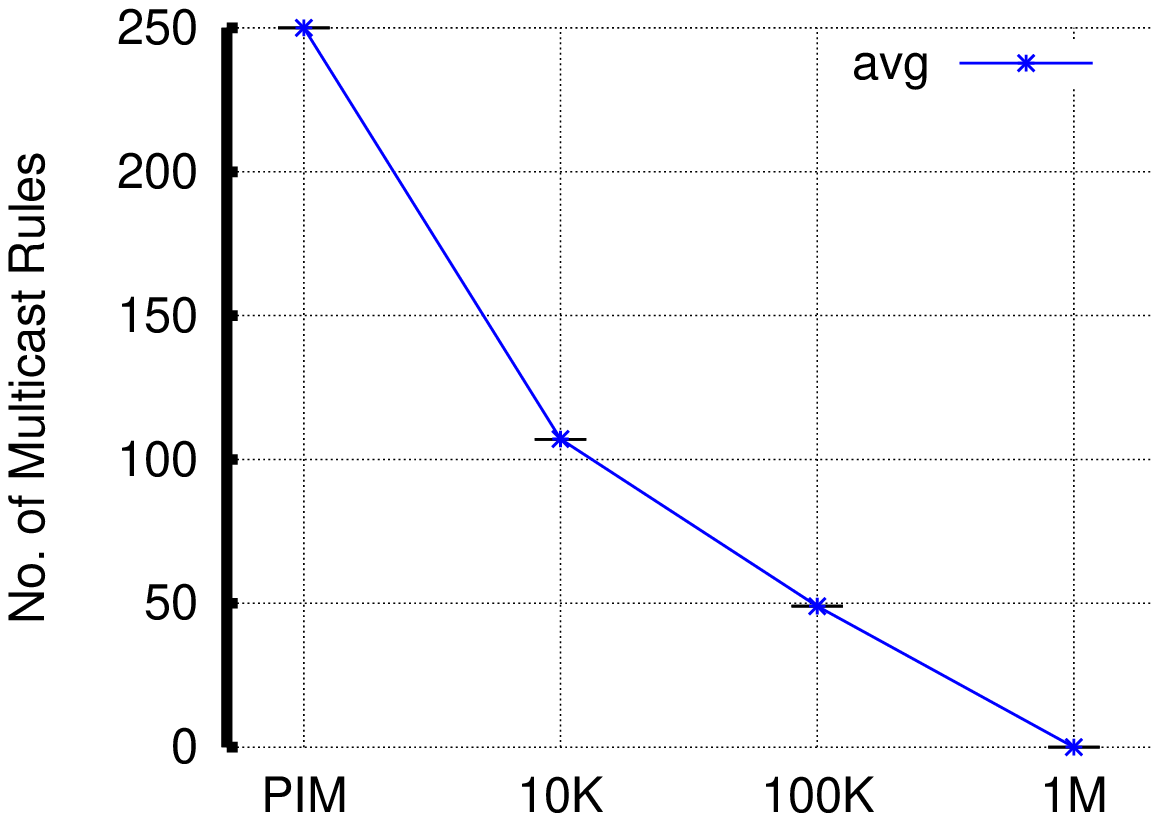}
   \label{fig:wve1024-16k-random-states-cores}
 }
\end{tabular}
\vspace{-1ex} \caption{Multicast state overhead of 16K groups on a 1024-server fat tree with Random placement } \label{fig:wve1024-16k-random-states}
\end{center}
\vspace{-4ex}
\end{figure*}

\begin{figure*}[t]
\begin{center}
\begin{tabular}{p{150pt}p{150pt}p{150pt}}
\subfigure[Edge State Capacity]
 {
   \includegraphics[width=5.7cm]{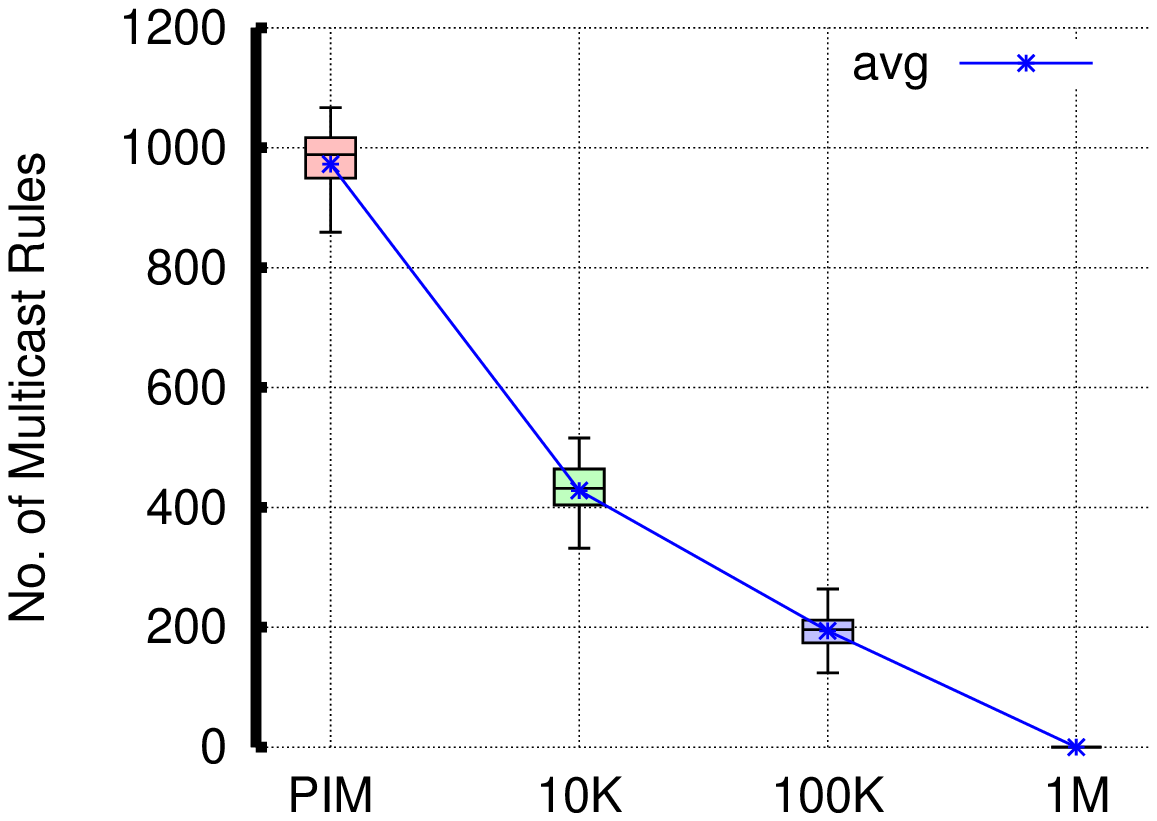}
   \label{fig:wve1024-64k-nearby-states-edges}
 }
 &
\subfigure[Aggr State Capacity]
 {
   \includegraphics[width=5.7cm]{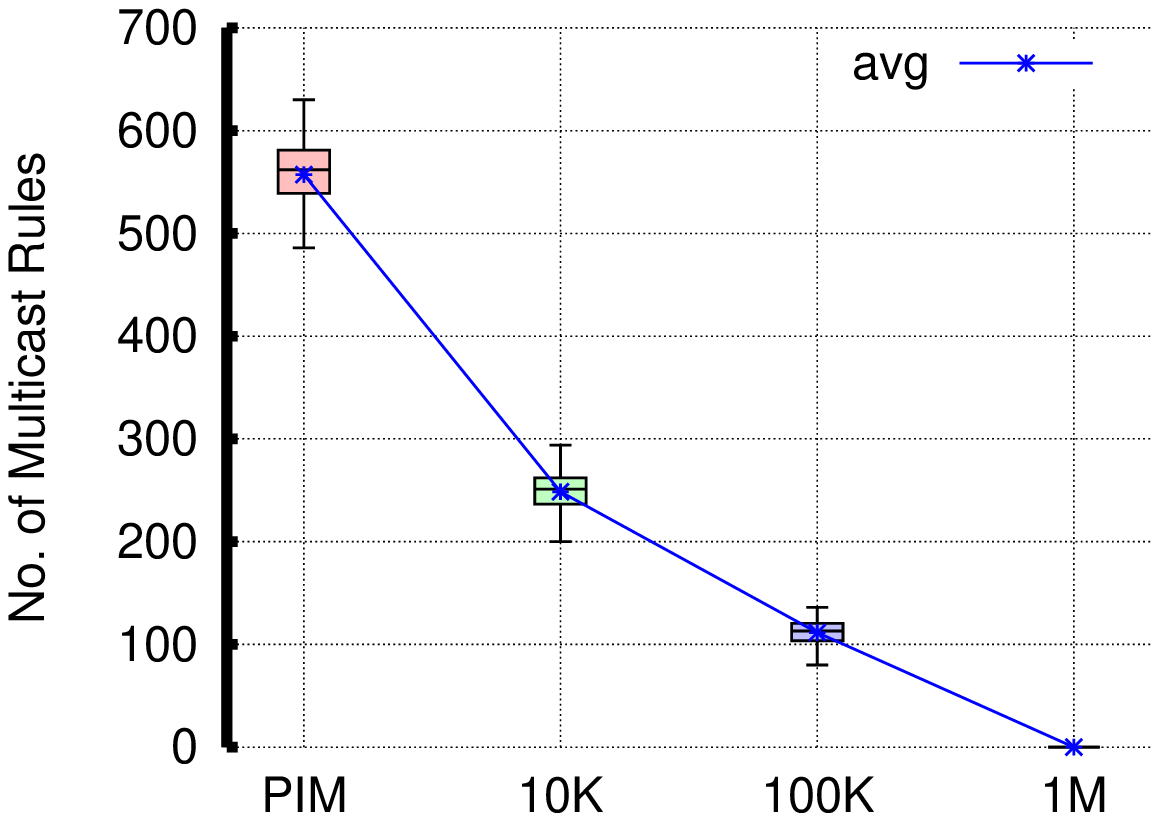}
   \label{fig:wve1024-64k-nearby-states-aggrs}
 }
 &
\subfigure[Core State Capacity]
 {
   \includegraphics[width=5.7cm]{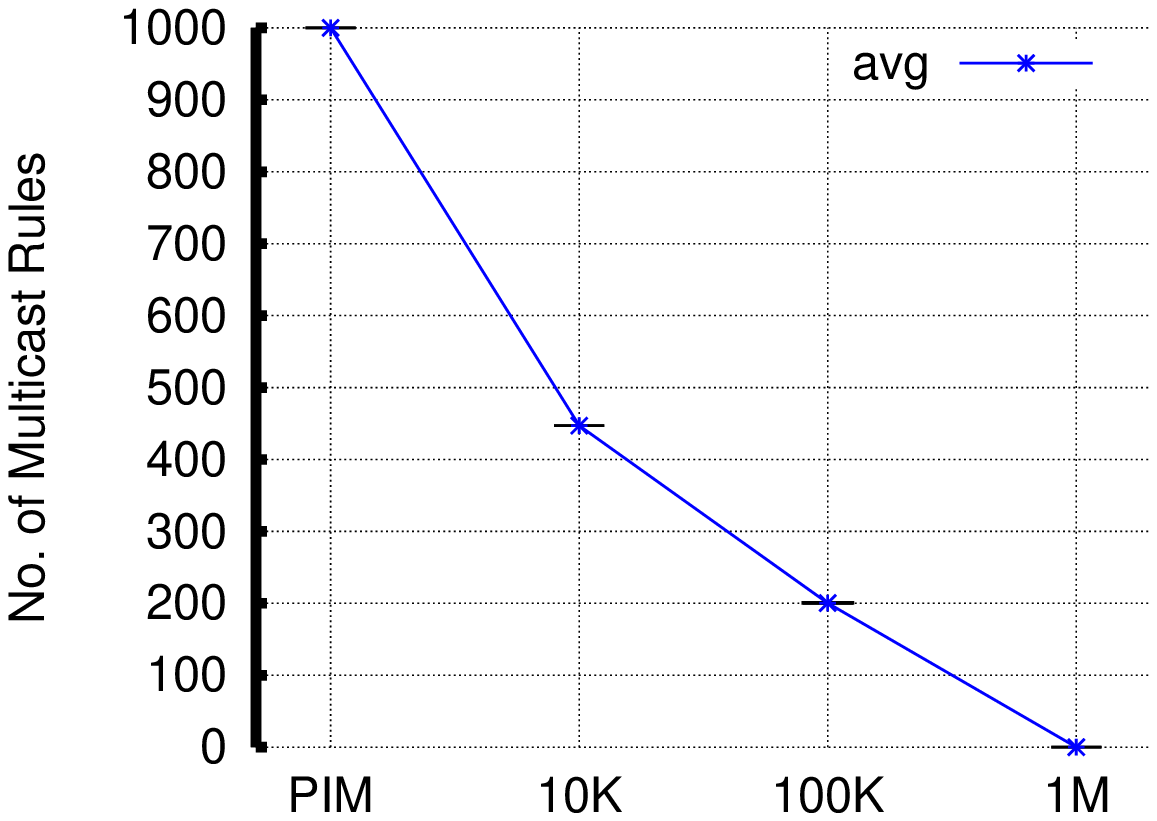}
   \label{fig:wve1024-64k-nearby-states-cores}
 }
\end{tabular}
\vspace{-1ex} \caption{Multicast state overhead of 64K groups on a 1024-server fat tree with Nearby placement} \label{fig:wve1024-64k-nearby-states}
\end{center}
\vspace{-4ex}
\end{figure*}

\section{Evaluation}
\label{sec:simulation}

\subsection{Methodology}
We evaluate the performance of DuSM on fat tree topologies. Since the IBM WVE and Telecom traces only have limited traffic data. We manipulate the real data by enlarging the traffic sizes and group numbers, while still keep the traffic distribution among groups.
We implement the DuSM system, including operation modules on switches, hypervisors, and the controller on the ns-2 simulator.
When mapping a group member (VM) to a physical server, we consider two different VMs placement strategies: (i) Random: group members are placed randomly across the network and the traffic has no locality, and (ii) Nearby: members of a same group are placed on nearby servers, reflecting some level of locality.

In our experiments, we vary the threshold used to separate mice and elephant groups by using three different values: 10 KB, 100 KB and 1 MB. 
When the threshold is set to 1 MB, DuSM treats most groups as mice. We compare DuSM with PIM-SM \cite{PIMSM}.

We focus on three performance metrics: multicast state overhead, balance of network traffic among different links, and number of switch updates under group dynamics.

\subsection{Multicast State Overhead}

Figure~\ref{fig:wve1024-16k-random-states} shows the number of  multicast rules of 16K groups on edge, aggregate, and core switches of a 1024-server fat tree with Random  placement, where \textbf{whiskers represent the min and max value; edges of the boxes
show the 25th, 50th (median), and 75th percentiles; and stars show the average values}. In general, the multicast state overhead shrinks as the threshold gets larger. When the threshold is 10 KB and 100 KB, our algorithm can save around 60\% and 80\% rules respectively. We find that the state capacity bottleneck is on the edge switches. Assuming the multicast state capacity of a commodity switch is 1000, PIM-SM can support around 15K groups, consistent to the results in \cite{Li:2013:SIM:2535372.2535380}. Using the 10 KB and 100 KB thresholds,  DuSM can support around 32K and 64K groups respectively.

Figure~\ref{fig:wve1024-64k-nearby-states} shows the number of  multicast rules of 64K groups on edge, aggregate, and core switches in a 1024-server fat tree with Nearby  placement. Similarly DuSM can significantly reduce memory cost on  switches.  Nearby placement allows the network to support more multicast groups due to traffic locality. Assuming the multicast capacity is 1000, PIM-SM can support less than 60K groups. Using the 10 KB and 100 KB thresholds,  DuSM can support more than 120K and 200K groups respectively.


\subsection{Network traffic distribution}
\begin{figure}[t]
\begin{center}
   \includegraphics[width=8cm]{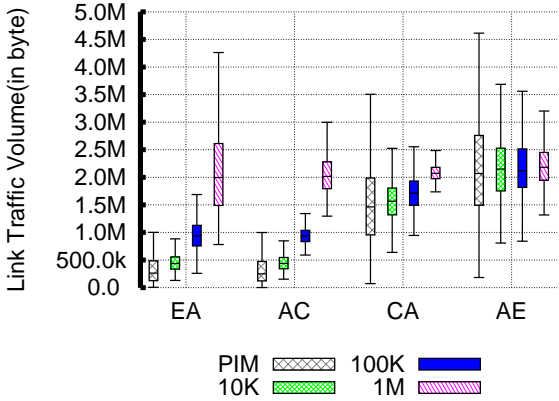}
   \label{fig:wve1024-16k-random-throughput}
%
\vspace{-3ex}
\caption{Traffic rate distribution of 16K groups on a 1024-server fat tree} \label{fig:wve1024-volume}
\end{center}
\vspace{-4ex}
\end{figure}

Figure~\ref{fig:wve1024-volume} shows the traffic distribution of 16K groups on different links of a 1024-host fat tree with Random placement. Here we consider four types of links: Edge to Aggregate (EA), Aggregate to Core (AC), Core  to Aggregate (CA), and Aggregate to Edge (AE). Note that multicast traffic on EA and AC links includes no replicate copies because EA and AC links are on the path from a sender to the root of the multicast tree. Traffic on CA and AE includes replicate copies to different receivers, and hence is higher than that on EA and AC. Therefore CA and AE links are easier to become bottlenecks. When DuSM is used, extra traffic overhead is introduced by the multicast/unicast translation of mice groups.  However DuSM (10 KB and 100 KB) only increases the average traffic on CA and AE links very little (less than 10\%).
More importantly, we can find that the maximum traffic rate for CA and AE links decreases as the threshold gets larger. Compared to PIM, DuSM (10 KB and 100 KB) can achieve better traffic balance by assigning packets to multiple shared trees and reduce the occurrence of network congestion. 

\subsection{Group Membership Dynamics}
We evaluate the influence of group membership dynamics by measuring the frequency of switch updates triggered by join and leave events. As discussed in Section~\ref{sec:controller}, the centralized controller will be informed of the dynamic event first and then disseminates corresponding actions on related switches. Generally speaking, only forwarding rules for elephant groups will be installed or removed. Table~\ref{tab:wve1024-16k-random-dynamic} and Table~\ref{tab:wve1024-16k-nearby-dynamic} show the switch updates for 16K groups on a 1024-server fat tree with different placement strategies. In general, we can conclude that DuSM can save around 90\% switch updates when we set the threshold to 10K, which is proportionate to the saved state capacity in Figure~\ref{fig:wve1024-16k-random-states}.

\begin{table}[t]
 \centering
 \begin{tabular}{|c|c|c|c|}
 \hline
  \# & PIM & 10K & 100K  \\ \hline  \hline
  Edge updates & 2096.66 & 207.98 & 85.24 \\ \hline
  Aggr updates & 1787.91 & 185.20 & 77.96  \\ \hline
  Core updates & 2483.64 & 259.52 & 109.52  \\ \hline
  Overall aver updates & 2050.56 & 406.67 & 169.02  \\ \hline
 \end{tabular}
\caption{Aver. no. of switch updates (Random)}\label{tab:wve1024-16k-random-dynamic}
\end{table}
\begin{table}[t]
 \centering
 \begin{tabular}{|c|c|c|c|}
 \hline
  \# & PIM & 10K & 100K  \\ \hline  \hline
  Edge updates  & 2111.67 & 198.45 & 81.97 \\ \hline
  Aggr updates  & 598.97 & 540.80 & 225.33  \\ \hline
  Core updates & 793.66 & 554.81 & 230.50  \\ \hline
  Overall aver updates  & 1242.98 & 181.61 & 79.22  \\ \hline
 \end{tabular}
\caption{Aver. no. of switch updates (Nearby)}\label{tab:wve1024-16k-nearby-dynamic}
\end{table}

\subsection{Summary of Evaluation}
In this section, we show that DuSM can support much more multicast groups than traditional IP multicast. The overall traffic volume does increase but the maximum traffic overhead reduces, indicating better load balance among links. In a hierarchical-tree topology with multiple paralleled paths, network congestion is usually caused by traffic imbalance rather than the overall traffic volume \cite{Hedera}. Hence DuSM does not hurt network scalability on bandwidth. DuSM also reduces the controller overhead for switch updates, which could be another potential scalability bottleneck \cite{Li:2013:SIM:2535372.2535380}. DuSM requires computing and memory resource on the hypervisor to perform multicast-to-unicast translation. We leave the evaluation of this cost to future work.

\section{Challenges and Open Issues}
\label{sec:challenges}
In this section, we will highlight some challenges and open issues of the complete design and implementation of DuSM.

\textbf{Dynamic group classification.}
Our evaluation exhibits a trade-off between multicast state  and traffic overhead. However, a proper value of the threshold for group classification usually relies on real network properties and communication environment. In the current design we assume a pre-defined threshold given the multicast traffic distribution is stable and known. In practice, we may prefer a dynamic tuning algorithm for group classification.

\textbf{Extension to general topologies.}
In this paper our design focuses on fat tree networks. Although DuSM can be easily extended to other hierarchical networks, we are also interested in whether we can apply this system to general topologies. In fact the major problem for elephant groups is to determine the multiple shared multicast trees in an arbitrary topology. The  RP (or core) selection problem  of DuSM is more complicated than core selection of IP multicast \cite{Calvert95core}, because multiple RPs are required to determine for an elephant group. In addition, the Steiner tree computation is NP-hard in an arbitrary topology. For mice groups, the problem is how to allow the unicast packets to avoid bottleneck links.

\textbf{Packet translation on switches.}
In this paper we suggest translating multicast packets to unicast packets  on hypervisors. If we move the translation module to edge switches, DuSM can save more bandwidth cost. Although some  current network techniques can achieve this function, such as IP over IP tunnelling \cite{rfc1853} and Generic Routing Encapsulation \cite{rfc1701}. A major concern is that how the edge switch knows the receivers of a multicast packet. One solution is writing all receiver addresses into the option field of an IP packet.

\textbf{Prototype implementation.} We will also implement a system prototype of DuSM using an OpenFlow-enabled platform.

\section{Related Works}
\label{sec:related}
Various multicast protocols has been proposed over the past two decades in the aspects of  reliable packet delivery \cite{Floyd:1997:RMF:270856.270863}, security \cite{judge03survey} and congestion control \cite{Widmer:2001:EEC:964723.383081} and scalability \cite{Jokela:2009:LLS:1594977.1592592, Vigfusson:2010:DMR:1755913.1755949, DanLi2011ESM, Cao:2012:DSE:2413176.2413182, Li:2013:SIM:2535372.2535380}. In this section, we restrict our consideration to scalable multicast protocols in Data Center Networks.

Jokela \emph{et al.} propose to encode multicast tree information into in-packet bloom filters and hence eliminate the use of multicast states \cite{Jokela:2009:LLS:1594977.1592592}. ESM \cite{DanLi2011ESM} enhances this approach by implementing source routing in DCNs with switch-based Bloom Filters.
Some other protocols improve scalability by locally aggregating multicast states at bottleneck switches \cite{Li:2013:SIM:2535372.2535380} or globally merging similar multicast groups \cite{Vigfusson:2010:DMR:1755913.1755949}. DataCast \cite{Cao:2012:DSE:2413176.2413182} leverages in-network packet caching and multiple Steiner trees in a DCN to address the scalability issues of IP multicast.

Unicast routing is widely adopted as a complementary technique for multicast protocols. Narada \cite{Chu:2000:CES:339331.339337} organizes end systems into an overlay structure and runs a distance vector protocol on top of the overlay for end-system multicast. Banerjee \emph{et al.} \cite{Banerjee:2002:SAL:633025.633045} design a scalable application-layer multicast solution using hierarchical clustering of multicast peers to avoid unnecessary packet replication. Dr.~Multicast \cite{Vigfusson:2010:DMR:1755913.1755949} uses a combination of point-to-point unicast and traditional IP Multicast by overloading the relevant socket operations.

\section{Conclusions}
\label{sec:conclusion}
Software defined networking enables the separation of multicast group management and forwarding. This paper proposes a dual-structure multicast system DuSM by utilizing the heterogeneity property of multicast traffic and centralized control for data center networks. DuSM uses state-free multicast  for  mice groups,  and uses multiple shared Steiner tree for a small number of elephant groups to balance their traffic in the network. Experimental results show that DuSM has great potential as a data center multicast system: it can support much more groups using commodity switches and achieves better load balance, compared to IP multicast. 

%
{\small
\bibliographystyle{abbrv}
\bibliography{pap}}  
%
%
\end{document}